\documentclass[structabstract]{aa}  
\usepackage{graphicx}
\usepackage{txfonts}
\usepackage{longtable}
\begin{document}
   \title{The galactic unclassified B[e] star HD\,50138\thanks{Based on observations made with the ESO Very Large Telescope Interferometer at Paranal Observatory (Chile) under the Guarantied Time Observing, programs 080.C.0089, 080.D.0181 and 082.D.0047, and from the ESO archive.}\thanks{This research made use of the Jean-Marie Mariotti Center \texttt{LITpro} service co-developped by CRAL, LAOG and FIZEAU.}}
   \subtitle{II. Interferometric constraints on the close circumstellar environment}

   \author{M. Borges Fernandes
          \inst{1,2},
	  A. Meilland
          \inst{3},
          P. Bendjoya
          \inst{2},
	  A. Domiciano de Souza
          \inst{2},
	  G. Niccolini
          \inst{2},
	  O. Chesneau
          \inst{2},
          F. Millour
          \inst{3},
	  A. Spang
          \inst{2},
          P. Stee
          \inst{2},
          \and
          M. Kraus
          \inst{4}
          }

   \institute{Observat\'orio Nacional, Rua General Jos\'e Cristino, 77, 20921-400, S\~ao Cristov\~ao, Rio de Janeiro, Brazil\\
	      \email{borges@on.br}
         \and
             Laboratoire H. Fizeau, UMR CNRS 6525, Universit\'e de Nice-Sophia Antipolis (UNS), Observatoire de la C\^ôte d'Azur (OCA), Campus Valrose, 06108, Nice, Cedex 2, France\\
             \email{olivier.chesneau@obs-azur.fr, philippe.stee@obs-azur.fr, philippe.bendjoya@unice.fr, armando.domiciano@unice.fr, gilles.niccolini@unice.fr}
         \and
             Max-Planck-Institut f\"ur Radioastronomie, Auf dem H\"ugel 69, 53121 Bonn, Germany\\
             \email{meilland@mpifr-bonn.mpg.de, millour@mpifr-bonn.mpg.de, } 
         \and
             Astronomick\'y \'ustav, Akademie v\v{e}d \v{C}esk\'e republiky, Fri\v{c}ova 298, 251\,65 Ond\v{r}ejov, Czech Republic\\
             \email{kraus@sunstel.asu.cas.cz}
             }

   \date{Received; accepted}

\authorrunning{Borges Fernandes et al.}
\titlerunning{The galactic unclassified B[e] star HD\,50138}

 
  \abstract
   {HD\,50138 is a southern star that presents the B[e] phenomenon, but its evolutionary stage is still not well known. This object presents spectral variability, which can be explained by outbursts or shell phases. Spectropolarimetric observations have shown the presence of a non-spherically symmetric circumstellar environment that is responsible for the B[e] phenomenon. However, up to now, the structure of the circumstellar medium of this object has not been studied deeply.    }
   { Based on recent optical long baseline interferometric observations from the VLTI/MIDI and VLTI/AMBER, and also from the Keck segment-tilting experiment, we study the structure of the circumstellar environment of HD\,50138.    }
   { The analysis of our data is based on geometrical analytical modeling, also using the recent LITpro software and considering a large space of parameters, which allows us to obtain a good estimate of the geometry of the circumstellar medium of HD\,50138, responsible for the emission in the H, K, and N-bands.   }
   { We resolve the circumstellar environment of HD\,50138 and describe its geometry for the first time in detail. Through analysis of multiwavelength data, the presence of a dusty circumstellar disk with an orientation onto the sky-plane of $71\pm7\degr$, which is perpendicular to the polarimetric measurements from the literature, was derived. We also derived that HD\,50138 is seen under an intermediate angle related to the line of sight, $56\pm4\degr$. In addition, the structure of the disk and the flux contributions of the gas and dust components is discussed. }
   { Based on analysis of different sets of interferometric data, we describe the circumstellar disk whose geometric parameters were determined, allowing us to understand the geometry of the circumstellar material of this bright star with the B[e] phenomenon.}


   \maketitle
%

\section{Introduction}

The B[e] phenomenon in stars with quite different evolutionary stages (Lamers et al. \cite{Lamers}) is still a puzzle. Several questions, especially those related to the physical processes responsible for this phenomenon and associated to the role of binarity or even of stellar mergers, are still being debated. This happens because only a few objects have been studied deeply up to now and more than 50\% of the B[e] stars have badly known or unknown evolutionary stages. 
 
In this paper, we continue the study of the southern galactic B[e] star \object{HD\,50138} (V743 Mon, MWC158, IRAS 06491-0654), started by Borges Fernandes et al. (\cite{BF09}, hereafter Paper I). This star has been studied since the beginning of the last century (Merrill et al. \cite{Merrill1}), and several papers have described the presence of a strong spectral variability. This variability has been explained by the presence of outbursts and shell phases (Hutsem\'ekers \cite{Hutsemekers}; Andrillat \& Houziaux \cite{Andrillat}). Paper I has suggested that a new shell phase took place before 2007, based on the analysis of high-resolution spectroscopic data. The stellar parameters of this object were also derived, confirming that HD\,50138 is a B6-7 III-V star. 

Despite numerous papers on the topic, the evolutionary stage of HD\,50138 remains highly uncertain. It has usually been classified either as a pre main-sequence Herbig Ae/Be object or as a classical Be star (Paper I and references therein). This star has also been classified as {\it ``a transition object between a classical Be and a B[e] star"} (Jaschek et al. \cite{Jaschek2}) or simply as an unclassified B[e] star (Lamers et al. \cite{Lamers}). In addition, the possibility of binarity cannot be discarded (Cidale et al. \cite{Cidale}; Baines et al. \cite{Baines}), similar to what is found by Millour et al. (\cite{Millour09}) for the unclassified B[e] star \object{HD\,87643}. A deeper discussion about the nature of HD\,50138 can be seen in Sect. 4 of Paper I.

In Paper I, the color excess of HD\,50138 due to purely interstellar extinction, i.e. $E(B-V)=0.08$\,mag, was also derived based on the existence of diffuse interstellar bands (DIBs). Moreover, from the analysis of the observed color indices, it was noted that the total extinction in the direction of this star was higher than the purely interstellar one. Therefore, this extinction excess must come from the  circumstellar environment of HD\,50138. However, how gas and dust are distributed in the close vicinity of HD~50138 is still unknown.  

Measurements from polarimetry and spectropolarimetry have shown that the circumstellar material of HD\,50138 is not spherically distributed. Based on UV and optical spectropolarimetry and on the analysis of H$\alpha$ intensity, Bjorkman et al. (\cite{Bjorkman}) show that electron scattering in a geometrically thin disk seen almost edge-on would be the main mechanism to produce the intrinsic polarization detected. On the other hand, Bjorkman et al. (\cite{Bjorkman}) also suggest that the dust responsible for the strong IR excess of HD\,50138 (Allen \cite{Allen}) would be either distributed in an optically thin spherical envelope or formed by large grains ($\geq$ 1$\mu$m), not contributing to the intrinsic polarization measured. Paper I showed that an ionized circumstellar disk or a dusty optically thin envelope, composed mainly of small grains, could be responsible for the circumstellar extinction. However other scenarios, like a dusty disk, where the dust is not located in the line of sight, could not be discarded. 

Bjorkman et al. (\cite{Bjorkman}) find a polarization angle of 158$\degr$, which is quite similar to the mean value of 160$\pm$2.6$\degr$ found by Yudin \& Evans (\cite{Yudin}) from UBV polarimetric measurements and of 157.5$\pm$5.0$\degr$ by Oudmaijer \& Drew (\cite{Oudmaijer}) from spectropolarimetric observations across the H$\alpha$ region. In addition, Oudmaijer \& Drew (\cite{Oudmaijer}) also suggest there are two different kinematic regions, where H$\alpha$ would be formed: a rotating disk and also a more spatially extended component. Assuming a disk geometry, this polarization angle can be perpendicular or parallel to the major axis of the disk, depending on whether the circumstellar material is optically thin or optically thick, respectively (Angel \cite{Angel}; Brown \& McLean \cite{BM}).

Recently, the advent of the optical long baseline interferometry using large telescopes, especially the VLT 8-m and the Keck 10-m telescopes, has made it possible to resolve, and consequently describe, the gaseous and dusty circumstellar environment of the brightest stars with the B[e] phenomenon (Domiciano de Souza et al. \cite{Domiciano}; Monnier et al. \cite{Monnier}; Millour et al. \cite{Millour09}; Meilland et al. \cite{Meilland}). The high spatial resolution acquired, associated to the spectral dispersion provided by instruments like VLTI/MIDI and VLTI/AMBER makes it possible to obtain sizes, shapes, and orientations of the circumstellar material, as a function of wavelength in the near and mid-IR ranges.

In this paper, our study is therefore focused on the description of the structure of the circumstellar material of HD\,50138 by the analysis of recent interferometric data from VLTI/MIDI and VLTI/AMBER, and also from Keck segment-tilting experiment. 

The paper has the following structure. In Sect. 2 we present the observations and the data reduction processes. In Sect. 3, we present the reduced data and our analysis that considers geometrical models. In Sect. 4, we present the discussion of our results and in Sect. 5, a short conclusion.


\section{Observations and data reduction}\label{obs}

\begin{table*}[!t]
      \centering
       \caption{VLTI/AMBER and VLTI/MIDI observing logs. }
      \begin{tabular}{cccccccc}
        \hline
        & & & & \multicolumn{2}{l}{projected baseline} & \\
        Number & Date & Obs.Time & Stations & Length & PA & Calibrators & Ang. Diam. of the Calibrators \\
        & & (UT) & &[meters] & [degrees] & (used in the triplet) & (mas) \\
        \hline
        \hline
        \multicolumn{8}{c}{AMBER (ATs)}\\
        \hline  
        B$_1$ & 2007-12-29 & 06:40 & G1-D0 & 62.3 & -36.4 & HD\,52436, HD\,52938, HD\,96538 & 1.24, 0.90, 0.92 \\ 
        B$_2$ &            &       & D0-H0 & 60.6 & 74.2 & \\
        B$_3$ &            &       & G1-H0 & 70.1 & 17.7 & \\
        B$_4$ & 2008-01-02 & 07:08 & K0-G1 & 90.3 & 35.2 & HD\,34137, HD\,44621, HD\,52938 & 0.83, 0.92, 0.90 \\
        B$_5$ &            &       & G1-A0 & 70.3 & -51.9 & \\
        B$_6$ &            &       & K0-A0 & 111.6 & 74.2 & \\
        B$_7$ & 2008-01-06 & 07:36 & G1-A0 & 62.3 & -44.5 & HD\,52938 & 0.90 \\
        B$_8$ & 2008-01-09 & 05:26 & H0-G0 & 31.4 & 73.8 & HD\,52938, HD\,54810 & 0.90, 1.49 \\
        B$_9$ &            &       & G0-E0 & 15.7 & 73.7 & \\
        B$_{10}$ & 2008-01-09 & 06:01 & H0-G0 & 30.2 & 74.2 & HD\,52938, HD\,54810 & 0.90, 1.49  \\ 
        B$_{11}$ &            &       & H0-E0 & 45.3 & 74.2 & \\
        B$_{12}$ & 2008-01-09 & 06:43 & H0-G0 & 27.9 & 74.2 & HD\,52938, HD\,54810 & 0.90, 1.49  \\
        B$_{13}$ &            &       & G0-E0 & 14.0 & 74.2 & \\
        B$_{14}$ &            &       & H0-E0 & 41.9 & 74.2 & \\
        B$_{15}$ & 2008-01-09 & 03:08 & H0-G0 & 29.7 & 68.3 & HD\,52938 & 0.90 \\
        B$_{16}$ &            &       & G0-E0 & 14.9 & 68.3 & \\
        B$_{17}$ &            &       & H0-E0 & 44.6 & 68.3 & \\
        \hline   
        \multicolumn{8}{c}{MIDI (ATs)}\\
        \hline   
        B$_1$ & 2007-12-09 & 03:19 & G1-H0 & 68.9 & -12.3 & HD\,48915 & 5.85  \\
        B$_2$ & 2007-12-09 & 03:40 & G1-H0 & 68.6 & -10.2 & HD\,48915 & 5.85 \\
        B$_3$ & 2007-12-12 & 05:09 & G1-D0 & 71.6 & 130.0 & HD\,48915 & 5.85 \\
        B$_4$ & 2007-12-12 & 06:00 & G1-H0 & 68.1 & 6.8 & HD\,48915 & 5.85 \\
        B$_5$ & 2007-12-13 & 03:10 & G1-D0 & 66.6 & 129.9 & HD\,48915, HD\,29139 & 5.85, 19.90 \\  
        B$_6$ & 2007-12-26 & 07:45 & G1-D0 & 58.1 & 151.3 & HD\,48915 & 5.85 \\
        B$_7$ & 2008-11-09 & 05:48 & E0-G0 & 12.3 & 60.7 & HD\,48915 & 5.85 \\
        B$_8$ & 2008-11-10 & 05:47 & E0-H0 & 37.1 & 60.9 & HD\,48915 & 5.85 \\
        B$_9$ & 2008-12-27 & 06:35 & E0-G0 & 15.5 & 74.0 & HD\,48915 & 5.85 \\
        B$_{10}$ & 2008-12-28 & 01:39 & E0-G0 & 10.0 & 50.8 & HD\,48915 & 5.85 \\
        B$_{11}$ & 2008-12-28 & 02:36 & E0-H0 & 36.9 & 60.7 & HD\,48915 & 5.85 \\
        B$_{12}$ & 2008-12-30 & 01:35 & G0-H0 & 20.3 & 51.7 & HD\,48915 & 5.85 \\
        B$_{13}$ & 2009-01-21 & 06:08 & G0-H0 & 27.1 & 74.0 & HD\,48915 & 5.85 \\
        B$_{14}$ & 2009-03-08 & 01:22 & E0-H0 & 47.7 & 73.4 & HD\,48915 & 5.85 \\
        \hline
      \end{tabular}
    \end{table*}
   
\begin{figure*}[!t]
\centering\includegraphics[width=0.8\textwidth]{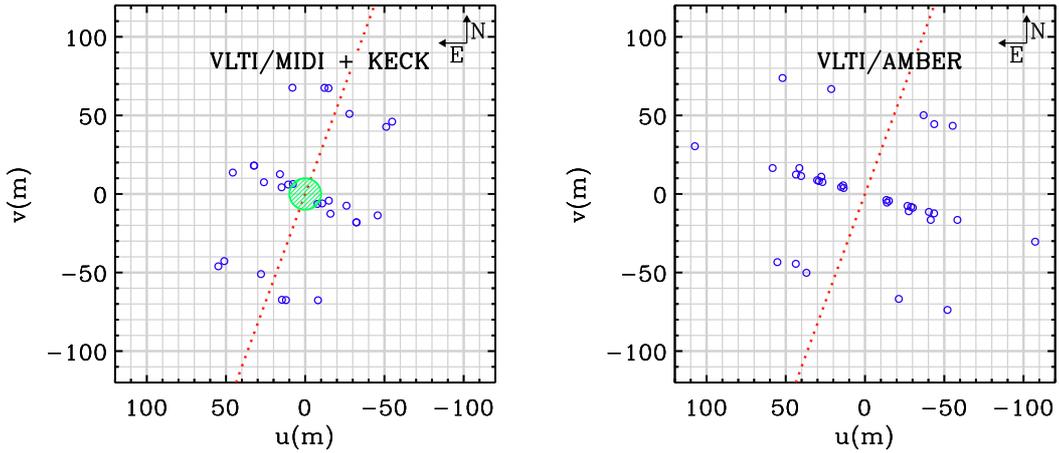}
\caption{The (u,v) plane coverage of HD\,50138 obtained with VLTI/MIDI (left panel) and VLTI/AMBER (right panel). The region covered by Keck data is also indicated as inside of the stripped circle in the left panel. The dotted line in both panels represents the polarization angle measured by Yudin \& Evans (\cite{Yudin}).} 
\label{uv}
\end{figure*}

\begin{figure}[t]
\centering\includegraphics[width=0.49\textwidth]{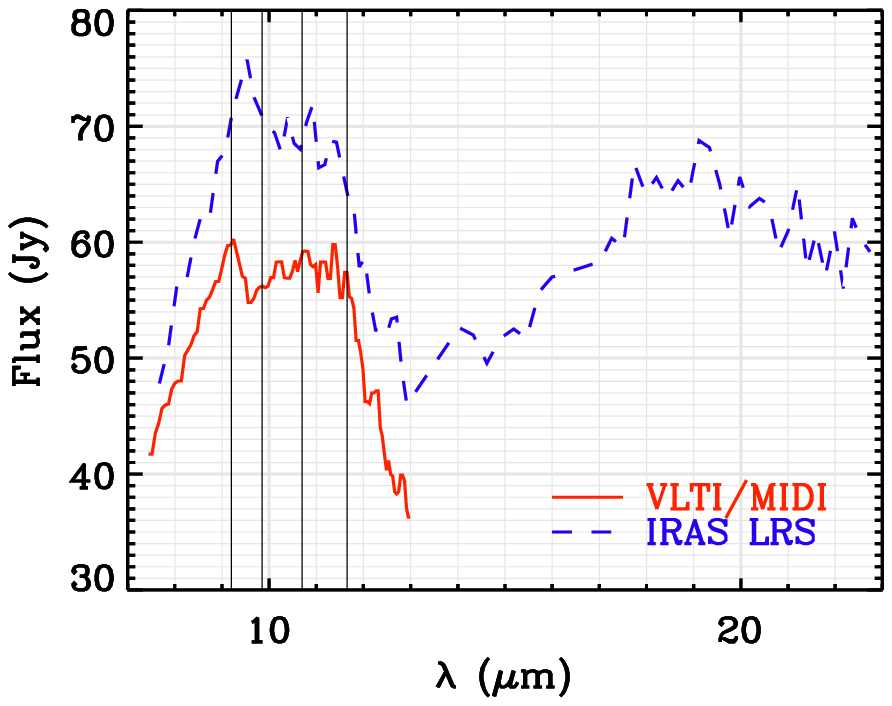}
\caption{VLTI/MIDI and IRAS LRS spectra of HD\,50138. The wavelengths of crystalline silicate grains in the VLTI/MIDI spectrum range, e.g. ortho-enstatite, are overplotted as vertical solid lines. } 
\label{MIDI}
\end{figure}

We obtained interferometric data on HD\,50138 using the VLTI MID-infrared Interferometric instrument (VLTI/MIDI, Leinert et al. \cite{Leinert}). The observations were carried out between December 2007 and March 2009 (guarantied time observing) using the 1.8-m auxiliary telescopes (ATs). The data were taken using the HIGH-SENS mode, where the photometry and the fringes are not recorded simultaneously. In total, 14 visibility measurements were obtained considering different projected baselines and orientations onto the sky-plane, as described in the Table 1. \object{Sirius} and \object{Aldebaran} were used as calibrators in these observing runs and their angular diameters come from MIDI Calibrator Catalogue\footnote{http://www.eso.org/sci/facilities/paranal/instruments/midi/tools/}. The (u,v) plane coverage of the VLTI/MIDI data of HD\,50138 is seen in Fig.~\ref{uv} (left panel). The data reduction was performed using two different packages: MIDI Interactive Analysis (MIA, Leinert et al. \cite{Leinert}) and Expert Work-Station (EWS, Jaffe \cite{Jaffe}), version 1.5.1. Since the two packages provide similar results within the error bars, we decided to use the EWS reduced visibilities in this work. VLTI/MIDI allows us to obtain not only spectrally dispersed fringes, but also the low-resolution spectrum (R = 30) of the source in the N-band region (7.5-13 $\mu$m). A comparison between VLTI/MIDI spectrum and IRAS low-resolution spectrum of HD\,50138 is shown in Fig.~\ref{MIDI}.

In addition to our VLTI/MIDI data, we also used 92 N-band (at 10.7 $\mu$m) visibility measurements obtained from the Keck segment-tilting experiment\footnote{Kindly provided by John Monnier}. It used 36 segments of 1.8-m each from the 10-m Keck telescope, forming four interferometric arrays, five subarrays, and baselines from 0.5-m to 9.0-m, by the tilting of the segments in the primary mirror (for more details, see Monnier et al. \cite{Monnier}). The indication of the region covered by Keck data in the (u,v) plane can be seen in the Fig.~\ref{uv} (left panel). 

We also analyzed sets of data at low spectral resolution (LR, i.e. R = 35) from five different observing nights in December 2007 and January 2008 from the Astronomical Multi BEam Recombiner, the near-infrared instrument of the VLTI (VLTI/AMBER, Petrov et al. \cite{Petrov}). We obtained data for two of these nights, and for the other three the data were taken from the ESO archive. A total of nine triplets, consequently measurements from 27 baselines, were available. However, based on the selection of the best quality data, we used only measurements from 17 baselines\footnote{The triplet taken on 2007-12-28, five baselines taken on 2008-01-06 and two on 2008-01-09 were not used. Thus only VLTI/AMBER data from four different nights were actually used by us.} (Table 1). The (u,v) plane coverage of the VLTI/AMBER data of HD\,50138 is presented in the Fig.~\ref{uv} (right panel). Measurements related to closure phases cannot be analyzed due to the large error bars presented (about 50$\degr$). The calibration stars were \object{HD\,52938}, \object{HD52436}, \object{HD96538}, \object{HD34137}, \object{HD44621}, \object{HD54810}, and \object{HD96068} (see Table 1), and their angular diameters come from Merand et al. (\cite{Merand}). The data reduction was performed with the standard VLTI/AMBER data reduction software (amdlib version 2.1, Millour et al. \cite{Millour04}; Tatulli et al. \cite{Tatulli}) and the frame selection was done as described by Cruzal\`ebes et al. (\cite{Pierre07}). 

\begin{table*}[!htb]
{\centering 
\caption{VLTI/MIDI visibilities and equivalent Gaussian FWHM.}
\begin{tabular}{c|ccc|ccc}
\hline 
   & \multicolumn{3}{c}{Visibility}\vline&   \multicolumn{3}{c}{Gaussian equivalent FWHM (mas)}\\
n$\degr$ & 8$\mu$m & 10.7$\mu$m & 13$\mu$m & 8$\mu$m & 10.7$\mu$m & 13$\mu$m\\
\hline
\multicolumn{7}{c}{Baselines roughly parallel to the Yudin \& Evans (2001) polarization measurement}\\
B$_{ 1}$& 0.10$\pm$0.06&0.16$\pm$0.07&0.30$\pm$0.12& 19.1$\pm$  2.7& 22.4$\pm$  2.8& 22.6$\pm$  4.1\\
B$_{ 2}$& 0.11$\pm$0.06&0.17$\pm$0.07&0.35$\pm$0.09& 19.1$\pm$  2.5& 22.4$\pm$  2.6& 21.1$\pm$  2.7\\
B$_{ 3}$& 0.07$\pm$0.05&0.09$\pm$0.05&0.22$\pm$0.06& 19.7$\pm$  3.4& 25.1$\pm$  3.4& 24.2$\pm$  2.4\\
B$_{ 4}$& 0.04$\pm$0.04&0.09$\pm$0.04&0.42$\pm$0.10& 22.6$\pm$  4.7& 25.9$\pm$  2.4& 19.5$\pm$  2.7\\
B$_{ 5}$& 0.09$\pm$0.06&0.10$\pm$0.06&0.21$\pm$0.09& 20.4$\pm$  3.6& 26.0$\pm$  3.3& 26.6$\pm$  3.8\\
B$_{ 6}$& 0.16$\pm$0.03&0.30$\pm$0.04&0.37$\pm$0.12& 20.2$\pm$  1.0& 21.7$\pm$  1.3& 24.4$\pm$  3.9\\
\hline
\multicolumn{7}{c}{Baselines roughly perpendicular to the Yudin \& Evans (2001) polarization measurement}\\
B$_{ 7}$& 0.61$\pm$0.05&0.62$\pm$0.05&0.78$\pm$0.16& 50.3$\pm$  4.3& 64.8$\pm$  5.7& 57.4$\pm$ 26.3\\
B$_{ 8}$& 0.11$\pm$0.05&0.14$\pm$0.06&0.35$\pm$0.12& 34.7$\pm$  3.7& 43.1$\pm$  4.5& 39.3$\pm$  6.6\\
B$_{ 9}$& 0.54$\pm$0.08&0.52$\pm$0.08&0.88$\pm$0.13& 44.2$\pm$  5.4& 60.1$\pm$  6.9& 33.5$\pm$ 24.9\\
B$_{10}$& 0.51$\pm$0.08&0.61$\pm$0.07&0.89$\pm$0.23& 71.8$\pm$  8.0& 80.3$\pm$  9.7& 49.4$\pm$ 46.4\\
B$_{11}$& 0.12$\pm$0.07&0.16$\pm$0.07&0.44$\pm$0.13& 34.2$\pm$  5.1& 42.1$\pm$  5.6& 34.8$\pm$  6.4\\
B$_{12}$& 0.53$\pm$0.10&0.56$\pm$0.10&0.75$\pm$0.16& 34.3$\pm$  5.2& 43.2$\pm$  6.5& 37.7$\pm$ 14.5\\
B$_{13}$& 0.30$\pm$0.04&0.37$\pm$0.06&0.55$\pm$0.08& 35.4$\pm$  2.2& 41.9$\pm$  3.2& 40.2$\pm$  5.1\\
B$_{14}$& 0.05$\pm$0.05&0.07$\pm$0.05&0.21$\pm$0.07& 32.1$\pm$  $>$11.8& 38.8$\pm$  6.1& 37.4$\pm$  4.2\\
\hline
\end{tabular}\par}
\label{baselines}
\end{table*}

\section{Geometrical models} 

The analysis of our interferometric data is based on geometrical analytical modeling, which has allowed us to obtain a description of the spatial distribution of the flux from the source in different bands (H, K, and N-bands), considering a large space of parameters. The first step was to analyse our N-band data obtained from Keck and VLTI/MIDI, which are mainly caused by circumstellar dust emission.

\subsection{N-band data (Keck+VLTI/MIDI)}

\subsubsection{The circumstellar environment at 10.7$\mu$m }

Monnier et al. (\cite{Monnier}) obtained data at 10.7 $\mu$m with the Keck segment-tilting experiment by considering different orientations and baseline lengths from 0.5 to 9.0\,m, making it possible to obtain information about the large-scale circumstellar environment. They could model these data by assuming both (i) a 1-D Gaussian disk (actually a round 2-D geometry) with a full width half maximum (FWHM) of 58$\pm$6\,mas, associated with an extended circumstellar envelope (called ``halo" by them) on scales larger than 0.5$\arcsec$, which would be responsible for only 1\% of the total flux in the N-band; and (ii) a 2-D Gaussian disk with FWHM for the major and minor axes of 66$\pm$4 and 46$\pm$9\,mas, respectively, also in association with a ``halo" responsible for just 1\% of the total flux in the N-band. The orientation of the major axis of the 2-D Gaussian disk was derived as 63$\pm$6$\degr$, which is roughly perpendicular to measurements of the polarization angle cited in the Sect. 1.

Thus, we have analyzed VLTI/MIDI data that have a broader spread of baselines in terms of lengths and orientations (Table 1), allowing us to constrain both the radial profile of the object and its flattening. These data can be divided into two groups in terms of baseline orientations (Table 2): from B$_1$ to B$_6$, which are roughly parallel to the polarization angle, and from B$_7$ to B$_{14}$, which are roughly perpendicular to it. The parallel baselines show an environment that is less resolved than seen by the perpendicular ones. This is shown in Table 2 from the estimate of the object extension (FWHM) assuming a simple 1-D Gaussian disk model. This result agrees with the orientation of the major axis of the disk found by Monnier et al. (\cite{Monnier}). 

\begin{figure*}[!tbh]
\centering   
\includegraphics[width=0.9\textwidth]{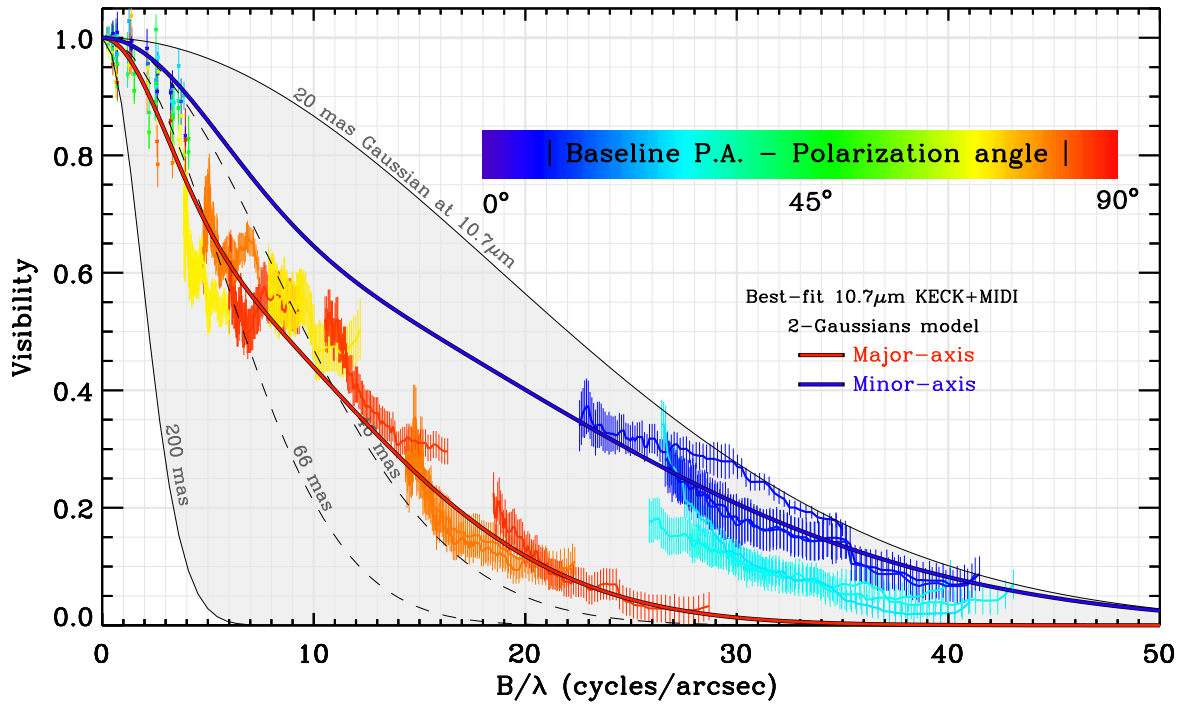}
\caption{VLTI/MIDI N-band spectrally resolved visibilities plotted as a function of the spatial frequency (B/$\lambda$) for the 14 baselines. Colors indicate the baselines position angle with respect to the polarization measurement by Yudin \& Evans (\cite{Yudin}). The color scale (i.e., blue, cyan, green, yellow, orange, and red) goes from blue for baselines parallel to the polarization angle to red for the perpendicular ones. The dots with error bars indicate the 10.7$\mu$m Keck measurements. The thin solid and dashed lines represent the visibilities from Gaussian disks with different FWHM. The thick red solid line represents the model fitting aligned to the PA of the major axes of the 2-Gaussian distribution and the blue one to the PA of the minor axes. }
\label{spfvis}
\end{figure*}

However, from our two groups of baselines, we derived a mean value for the FWHM of the major and minor axes at 10.7 $\mu$m of 52$\pm$15 and 24$\pm$2\,mas, respectively. The large difference compared to the results of Monnier et al. (\cite{Monnier}), i.e., a factor of 1.3 for the major and 1.9 for the minor axis, shows the limitation of their calculations owing to the absence of longer baselines. In Fig.\,\ref{spfvis}, we have plotted the VLTI/MIDI and Keck visibilities as a function of the spatial frequency, and we note that different single Gaussian disks with FWHM from 20 to 100\,mas cannot model them. Since the trend in the visibilities is not compatible with other simple models, such as uniform disk or ring, we need to find a geometric analytical model that simultaneously fits the low spatial frequencies probed by the Keck and the higher ones measured by the VLTI/MIDI. We test two different models that are compatible with this qualitative analysis:

\begin{enumerate}
\item A model consisting of two elliptical Gaussian distributions (representing a compact and a larger component) with the same flattening and orientation of the major axis ($\theta$). The other parameters are the FWHM of the major axis of each Gaussian and the relative flux of the compact component F$_1$ (i.e., FWHM of the second Gaussian, F$_2$ = 1 - F$_1$). Thus, this model has five free parameters. \\

\item A flattened exponential distribution, i.e. I(r)~$\propto$~e$^{-\left|r/r_0\right|}$. The model parameters are the flattening, the orientation of the major axis ($\theta$), and r$_0$. To enable direct comparison with the Gaussian distribution, we use the FWHM of the exponential distribution instead of r$_0$. A simple calculation using the formula I(FWHM/2) = 0.5, directly gives FWHM = 1.38 r$_0$. This model only has three free parameters. 

\end{enumerate}

Finally, we also use a single elliptical Gaussian distribution as a reference model. To not be biased by a putative change in the envelope geometry between 8$\mu$m and 13$\mu$m, we decided to first focus on the visibilities at 10.7$\mu$m. Using the VLTI/MIDI and Keck data, we have a set of 106 measurements. The model fitting is done by minimizing the reduced chi-square ($\chi^2_{r}$) on the visibilities, using a Levenberg-Marquardt (LM) two-dimensional algorithm. As this technique allows converging to local minimum, we use 10$^4$ starting random positions in the parameter space to be sure that we reach the global minimum. The result of this model fitting is presented in Table~\ref{10.7model}.

As already noticed in Fig.\,\ref{spfvis}, a single elliptical Gaussian distribution is not able to fit our dataset correctly, since its $\chi^2_r$ is significantly higher than for the other two models. The 2-Gaussian model slightly shows the best $\chi^2_r$, but the characteristic sizes, based on the FWHM of the major axis of the exponential distribution and of the compact component of the 2-Gaussian distributions, are compatible within {\bf 3$\sigma$}, i.e., 30.7$\pm$1.4 and 35.2$\pm$1.5 mas, respectively. Moreover, the flattening and orientation of the major-axes are consistent within 1$\sigma$. In addition, we could also estimate the inclination related to the line of sight of the circumstellar structure, based on the flattening and assuming the 2-Gaussian and exponential distributions, i.e., 56.7$\pm$0.4$\degr$ and 59.0$\pm$1.6$\degr$ respectively, which are also compatible within 2$\sigma$.

\begin{table}[!htb]
\caption{Parameters of the best-fit models from our 10.7$\mu$m analysis.}
{\centering \begin{tabular}{cccc}
\hline
 &Gaussian& 2-Gaussians & Exponential\\
\hline
F$_1$ & - & 0.68 $\pm$ 0.04 &-\\
FWHM$_{major1}$ (mas) & 64.7 $\pm$ 0.6  & 35.2 $\pm$ 1.5 & 30.7 $\pm$ 1.4\\
FWHM$_{major2}$ (mas) & - & 131.4 $\pm$ 11.2 & - \\
Flattening   &   2.94 $\pm$ 0.10 & 1.82 $\pm$ 0.02 & 1.94 $\pm$ 0.09\\
$\theta$ ($\degr$)& 59.1 $\pm$ 1.7 & 65.9 $\pm$ 2.0 & 64.1 $\pm$ 2.3\\
$i$ ($\degr$) & 70.1 $\pm$ 0.7 & 56.7 $\pm$ 0.4 & 59.0 $\pm$ 1.6 \\
\hline
$\chi^{2}_{r}$& 5.1 & 1.9& 2.4\\
\hline
\end{tabular}\par}
\label{10.7model} 
\end{table}

\subsubsection{Wavelength dependence through the N-Band}

The VLTI/MIDI data show an object that is partly resolved for all baselines with a slight increase in visibilities with the wavelength. This mainly stems from the wavelength dependence of the interferometer spatial resolution, i.e. $\phi_{min} \propto \lambda$/B, which is a factor 1.6 higher at 8$\mu$m than at 13$\mu$m, explaining the observed increase in visibility with the wavelength. No indication of a more extended envelope in the silicate band region is seen, unlike what is expected, since the opacity is supposed to be significantly higher in this band than in the continuum. However, as shown in Fig.~\ref{MIDI}, the silicate band is strongly in emission and is very broad, covering almost all the 8-13$\mu$m VLTI/MIDI spectral band. This could explain why no clear drop is seen in the visibilities at the center of the silicate band.

\begin{figure*}[tbh]
\centering\includegraphics{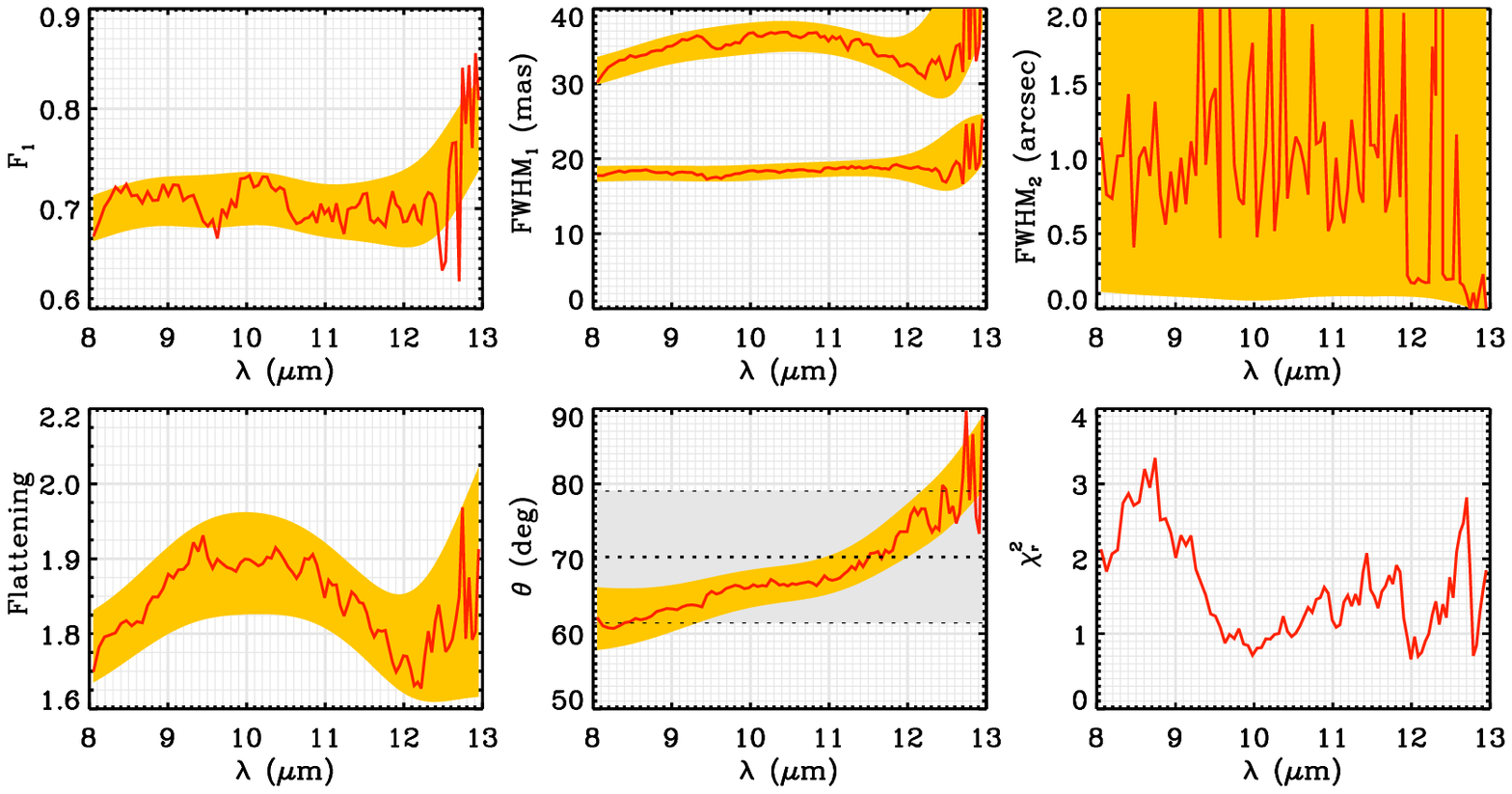}
\caption{Best-fit model parameters and $\chi^2_r$ for the 2-Gaussian model considering only the VLTI/MIDI data. All the parameters are plotted as a function of the wavelength: flux ratio of the compact component (top left); FWHM of the major and minor axes of the compact component (top center); FWHM of the major-axis of the large component (top right); flattening (bottom left); orientation of the major-axis (bottom center); and $\chi^2_r$ (bottom right).} 
\label{fit}
\end{figure*}

To quantitatively constrain the geometry of the circumstellar environment through the N-band, we used the same fitting method described in the previous section, considering only the 2-Gaussian model, which provides the best results. For each VLTI/MIDI spectral channel, we fit the visibility of the 14 baselines. Thus, the previously defined model parameters are now functions of the wavelength. In Fig.\,\ref{fit}, we plotted the value of the best-fitted parameters as a function of the wavelength for the 2-Gaussian model. 

We calculated an average $\chi^2_r$ over all spectral channels, i.e., $< \chi^2_r(\lambda) >_\lambda$, and found a value of 1.5. Moreover, as expected from the global trend of the spectrally resolved visibilities, the geometry does not strongly depend on the wavelength.  The FWHM of major axis of the compact component varies from 30$\pm$2~mas at 8\,$\mu$m to 37$\pm$3~mas around 10$\mu$m and return to 32$\pm$3~mas around 12\,$\mu$m. It remains unclear whether these small variations, on the order of 2$\sigma$, are significant or not. This is also the case for the flattening of the compact component with variations between 1.7 and 2.0, and for the orientation of the major axis with variations between 60$\degr$ and 80$\degr$, which are both on the order of 2$\sigma$. The relative flux of the compact component is stable through the whole N-band; i.e., F$_1$ = 0.7$\pm$0.03. Finally, the size of the extended component is larger than 100\,mas. This means that it is fully resolved even by the shorter baselines. 

By analyzing the results in the frame of a circumstellar dusty disk, we can constrain a few geometrical parameters of the disk. First, we can deduce that the inner radius of the dusty disk lies within the compact component, so that its extension is smaller or equal to 15\,mas. Considering a distance of 500\,pc, this corresponds to an inner radius of the dusty disk smaller than 7.5\,AU. Second, we can directly derive the inclination angle of the disk in the case of the geometrically thin disk; i.e., i = arccos (b$_1$/a$_1$) = 57$\pm$3$\degr$. However, the opening angle of the disk may not be negligible, so that we can, as shown by Meilland et al. (\cite{Meilland1}), only put a lower limit on the inclination of the object, i $\ge$ 54$\degr$. Finally, we can also confirm that the major axis of the disk is roughly perpendicular to the polarization measurements, $\theta$ = 70$\pm$12$\degr$.

\subsection{H and K-band data (VLTI/AMBER)}

The analysis of data from this spectral region is not trivial, since the near-IR emission usually comes from different sources: the stellar radiation, the hot gas located closer to the star than the dust sublimation radius (in a dust-free region), and also the hot dust, mainly located at the inner radius of the dusty disk.

\begin{figure*}[!t]
\centering\includegraphics[width=0.8\textwidth]{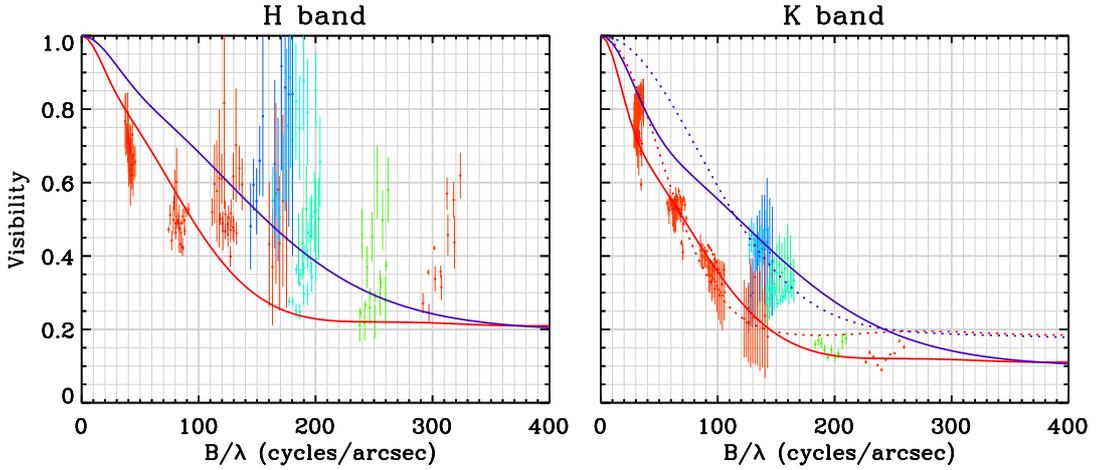}
\caption{Model fitting of the VLTI/AMBER data, using the parameters obtained by LITpro: right panel for the K-band assuming both scenarios as described in the text for Model 1 (dotted lines) and 2 (solid lines), and left panel for the H-band assuming a scenario like Model 2. The different colors follow the same description as in Fig.~\ref{spfvis}, representing the orientation of the baselines. } 
\label{amberKmodel2}
\end{figure*}

We started our analysis of the near-infrared VLTI/AMBER measurements by modeling the K-band data. The visibilities are plotted as a function of the spatial frequency in Fig~\ref{amberKmodel2} (right panel), where we clearly see that the object is fully resolved by the longest baselines. However, a non-negligible part of the flux, in the range of 10-20$\%$, is still unresolved, which probably represents the stellar flux. 

Another interesting point is the absence of a sharp edge, such as an inner rim of the dusty disk, which would produce oscillations in the visibilities. This absence can be for three different causes: (a) the inner radius is very extended and thus over-resolved even by the smallest baselines, (b) its contribution to the total flux is small enough to reduce the oscillation amplitude to an undetectable level, and (c) there is no gap in emission between the inner gaseous envelope/disk and the inner rim of the dusty disk. We note that a combination of these causes can also be possible. Whatever causes the absence of oscillations, it will strongly prevent us from disentangling the gas and dust emission in the K-band.

Thus, to perform a model fitting of these data, we decided to test LITpro\footnote{LITpro software available at http://www.jmmc.fr/litpro} (Lyons Interferometric Tool prototype), a new software developed by the Jean-Marie Mariotti Center (JMMC). It uses a set of elementary geometrical functions and also the LM algorithm to minimize the $\chi^2_r$. It is important to cite that we did not use LITpro for the N-band analysis, because it is not yet able to perform an achromatic model fitting. 

We decided to start our model fitting by using the simplest model that could fit the data, i.e. an unresolved star plus an elliptical Gaussian model. Such a model can help us to derive the characteristic extension of the circumstellar environment. This simple model (hereafter Model 1) has the following parameters: the stellar relative flux (F$_\star$), and the Gaussian relative flux (F$_1$), its FWHM of the minor axis (FWHM$_{minor1}$), the orientation of the major axis on the sky-plane ($\theta$), and the flattening. Since the flux is normalized, we have F$_1$ = 1 - F$_\star$, and consequently, Model 1 has four free parameters.

The result of the model fitting is presented in Table~\ref{amber}, and with $\chi^2_r$ = 40.8, we see that Model 1 failed to reproduce the spatial-frequency visibility variations (Fig.~\ref{amberKmodel2}, right panel). However, we note that the flattening and orientation of the model 1, i.e. 1.7$\pm$0.3 and 66$\pm$9$\degr$, are compatible within 1$\sigma$ with those ones derived from the 10.7$\mu$m modeling. In addition, we could also, based on the flattening, estimate the inclination related to the line of sight of the circumstellar structure, i = 54$\pm$8$\degr$, which also agrees with our N-band results.

Since the gas and dust emission regions may have different characteristic sizes, we also tried to model the environment with a more complex geometry, consisting of an unresolved star plus two Gaussian distributions, with the same flattening and orientation onto the sky plane. This model (hereafter Model 2) has the following parameters: the stellar flux (F$_\star$), the first Gaussian flux (F$_1$) and its FWHM of the minor axis (FWHM$_{minor1}$), the second Gaussian relative flux (F$_2$), its FWHM of the minor axis (FWHM$_{minor2}$), and the Gaussians flattening and orientation ($\theta$). As described for Model 1, owing to the normalized flux, Model 2 has six free-parameters. We obtained $\chi^2_r$ = 13.3, which is significantly better than the one for Model 1, but not yet fully satisfying. 

We also tried to use other intensity distributions, such as uniform disks, rings, and exponentials, but we did not manage to obtain a better fit. Thus, we decided to keep Model 2 as the reference for the K-band (see Table~\ref{amber} and Fig.~\ref{amberKmodel2}, right panel).

\begin{table}[!t]
\caption{Results of the K-band model-fitting using LITpro.}
{\centering \begin{tabular}{ccc}
\hline
 & \multicolumn{2}{c}{K-band} \\
 Parameter & model 1 & model 2\\
\hline\hline
F$_\star$ & 0.19 $\pm$ 0.03 & 0.12 $\pm$ 0.01 \\
F$_1$ & 0.81 $\pm$ 0.03 & 0.61 $\pm$ 0.06 \\
F$_2$ &                 & 0.27 $\pm$ 0.07 \\
flattening & 1.7 $\pm$ 0.3 & 1.7 $\pm$ 0.3 \\
$\theta$ ($\degr$) & 66 $\pm$ 9 & 77 $\pm$ 2 \\
FWHM$_{minor1}$ (mas) &   4.4 $\pm$ 0.5 & 3.0 $\pm$ 0.4 \\
FWHM$_{minor2}$ (mas) &  & $\ge$ 14.0 \\
$i$ ($\degr$) & 54 $\pm$ 8 & 54 $\pm$ 8 \\
\hline
$\chi^2_r$ & 40.8 & 13.3 \\
\hline
\end{tabular}\par}
\label{amber} 
\end{table}

The orientation of the envelope, its flattening, and its inclination related to the line of sight, determined by Model 2 (1.7$\pm$0.3, 77$\pm$2$\degr$, and 54$\pm$8$\degr$, respectively) are also compatible with our N-band analysis. The most compact Gaussian major axis has an extension of 5.1$\pm$1.1\,mas, corresponding to $\sim$ 2.6\,AU, whereas the extended one cannot be fully constraint, but it is larger than 24\,mas, i.e. 12\,AU. The relative flux of the compact component (61$\pm$6$\%$) dominates the K-band emission.

As stated before, it is really hard to conclude about the sizes of the gas and dust emission regions from this simple modeling. We know from the VLTI/MIDI modeling that the inner rim of the dusty disk is smaller than 7.5 AU, but both the compact (HWHM $\sim$ 1.3\,AU) and extended components (HWHM $\sim$ 6\,AU) are compatible with this. Thus, we have tried to substitute in our models one of the components by a ring. Unfortunately, we did not manage to obtain any better agreement with such a model than for Model 2. Nevertheless, the ring radius cannot be smaller than 3 AU, otherwise oscillations in the visibilities would be seen even for the longest baselines. Thus, it finally seems that the large component probably represents the inner radius of the dusty disk. 

In addition, we have also tried to model the H-band VLTI/AMBER data, where the quality is clearly lower compared to the K-band data (Fig.~\ref{amberKmodel2}, left panel). Using the same scenario as described by Model 2, we did not manage to constrain properly the model parameters, especially the extensions, flattening, and orientation of the circumstellar environment. However, we can consider that only the relative flux of the components would change in the H-band compared to the K-band. Thus, we fixed the geometrical parameters with the same values as used for the K-band, and we derived the fluxes for the three components: F$_\star$ = 0.22 $\pm$ 0.04, F$_1$ = 0.65 $\pm$ 0.10, and F$_2$ = 0.13 $\pm$ 0.07. The corresponding visibilities are overplotted in Fig.~\ref{amberKmodel2} (left panel).

\section{Discussion} 

A good coherence was obtained considering the model-fitting of VLTI/AMBER, VLTI/MIDI and Keck data, allowing us to resolve and describe in detail for the first time the circumstellar environment of HD\,50138. In addition, based on our data and their different project baselines and position angles, we cannot verify that there is variability in the interferometric measurements associated to the spectral variations of this object that were described in Paper I. Thus, we can now draw a general picture for this object.

\subsection{General structure of the dusty disk}

Based on our model fitting for different data, we saw that the star is seen under an intermediate inclination angle of 56$\pm$4$\degr$ and is surrounded by a large dusty disk, with a major-axis roughly perpendicular to the polarization angle, $\theta$ = 71$\pm$7$\degr$). This disk probably has a small opening angle, which can explain the photospheric lines seen in the high-resolution spectra of HD\,50138 (Paper I). Moreover in this case, since the dust would be out of our line of sight, it does not significantly contribute to the visible polarization, in agreement with the results of Bjorkman et al. (\cite{Bjorkman}). 

From our models in the mid-IR, we see that the dusty disk emits up to a large distance from the central star, i.e. more than 30 AU. Moreover, the flux measured by the VLTI/MIDI spectrum is about 13\% lower than the value provided by IRAS LRS (see Fig~\ref{MIDI}), both with similar spectral resolutions. This difference can be considered quite small, since the uncertainty of the VLTI/MIDI flux is around 20\%. In addition, the VLTI/MIDI field of view (FOV) is around 1$\arcsec$ at 8$\mu$m and of 1.5$\arcsec$ at 13$\mu$m, while the IRAS LRS one covered a region of 7.5$\arcmin$. Based on this, we can consider that VLTI/MIDI spectrum represents the whole N-band emission of HD\,50138, thereby avoiding any external contamination. 

\subsection{Mineralogy of HD~50138 dust}

The VLTI/MIDI spectrum also allows us to qualitatively analyze the mineralogy of the circumstellar medium of HD\,50138. We note the presence of a 10$\mu$m emission bump that corresponds to the silicate band, which does not seem to affect the visibility measurements. However, as already mentioned in Sect.\,3.1.2, this is probably because the silicate band is as broad as the VLTI/MIDI spectral bandwidth.

In addition, the VLTI/MIDI spectra, even from shorter baselines, show features from cristalline silicate grains, such as ortho-enstatite (Fig.~\ref{MIDI}). This does not agree with the expectations for young stellar objects (van Boeckel \cite{vanBoekel} and di Folco et al. \cite{diFolco}). However, a more detailed analysis of the dust composition, associated to a larger sample of data, becomes necessary to provide a complete study of the dust content, as a function of the distance from the star.

\subsection{Separating the stellar, gas, and dust emissions}   

In our K-band modeling, we found that HD\,50138 circumstellar environment can be divided into two components. The first one, fully resolved even by the smallest baselines, contributes to 27$\pm$7$\%$ of the flux. A lower limit on its extension around 24\,mas (12\,AU) can be set. Comparing this component with the N-band modeling, we conclude that it probably represents the K-band contribution of dusty disk inner rim. However, only a radiative transfer code, which couples gas and dust, will allow us to derive the radius of the inner dusty disk properly. It is also clear that, even in this case, we will not find an specific inner radius, but a range of radii, depending of the grain size. 

The second component is much more compact with FWHM = 5.1$\pm$1.1\,mas (i.e. 2.6\,AU), and it represents 61$\pm$6$\%$ of the K band, which can be associated to hot gas emission. Concerning the mid-IR emission, since the N-band stellar flux itself is obviously negligible, it can only originate in the dust surrounding HD\,50138.

Through the stellar, dust, and circumstellar gas relative fluxes, and photometric and spectro-photometric measurements in H, K, and N bands, it is possible to derive the flux contribution of each component. We used 2MASS measurements to estimate the total flux for the H and K-bands, and the VLTI/MIDI spectrum (as cited in the Sect. 4.1) for the total flux of the N-band. The resulting fluxes are shown in Table~\ref{tableflux}.
 
\begin{table}[!b]
\caption{Fluxes of the components of HD\,50138 for the different bands.}
{\centering \begin{tabular}{cccccc}     
\hline
 Band &$\lambda$&\multicolumn{4}{c}{Fluxes in W m$^{-2}$ $\mu$m$^{-1}$}\\
 &($\mu$m)& Star 					& Gas						 	& Dust 						& Total \\
 \hline\hline
 H &1.65							&	2.4 10$^{-12}$&	7.0 10$^{-12}$	&	1.4 10$^{-12}$	&1.1 10$^{-11}$\\
 K &2.20							&	1.1	10$^{-12}$&	5.6 10$^{-12}$	&	2.5 10$^{-12}$	&9.2 10$^{-12}$\\
 N &8									&		-						&	   -               &	2.3 10$^{-12}$	&2.3 10$^{-12}$\\
 N &10.7							&		-						&	  	   -       &	1.6 10$^{-12}$	&1.6 10$^{-12}$\\
 N &13								&		-						&	-	          &	6.4 10$^{-13}$	&6.4 10$^{-13}$\\
\hline
\end{tabular}\par}
\label{tableflux}
\end{table}

\subsection{The gaseous inner disk}

Thanks to the previous section, we now have suggested that the compact component of our K-band model represents the emission of the hot gas surrounding HD\,50138. Moreover, the flattening of this component, i.e. 1.7$\pm$0.3, is compatible with the one determined in N-band for the dusty disk, i.e. 1.82$\pm$0.02, considering the geometrical model at 10.7$\mu$m. The major axis orientation of these two components is also compatible. Therefore, most of the hot gas must be located in a disk. This disk may be a Keplerian accretion disk, but just in case of a young nature for HD\,50138, or even an outflowing disk wind for a more evolved scenario. 

To distinguish between these different scenarios, it is necessary to constrain not only the circumstellar environment geometry but also its kinematics. This can be done by coupling the high-angular resolution of modern interferometers with a sufficiently high spectral resolution. Based on this, observing programs for VLTI/AMBER are currently being prepared, using its high spectral resolution mode (R=12000).

Another interesting point is to compare the assumption of a geometrically thin disk seen under an intermediate inclination angle with the Balmer lines profiles described in Paper I, i.e. H$\alpha$, H$\beta$, H$\gamma$, and H$\delta$. In spite of their variations, all these emission lines clearly show a strong and narrow absorption component that is probably formed in the circumstellar environment. Thus, a non-negligible part of the emitting gas must be located in the line of sight, i.e 56$\degr$, hence far above the equatorial disk. Consequently, this gaseous emission could come from the stellar wind, as described for Herbig stars by Kraus et al. (\cite{Kraus2008}).

\section{Conclusions} 

HD\,50138 is a curious star whose evolutionary stage is unknown. In this paper, we could describe the geometry of both the gaseous and dusty circumstellar environments, based on the analysis of recent interferometric data. Through geometrical analytical modeling, it was possible to resolve a circumstellar disk around HD\,50138 and to describe its parameters for the first time in detail.
 
However, more observations, especially from the VLTI and with high S/N, are planned. The analysis of these data, in association with 3-D codes, will allow us to derive not only the physical parameters of the dusty environment precisely, but also those from the gas emitting regions. In addition, new data will make the image reconstruction of HD\,50138 possible. This information will be extremely valuable for better understanding the physical mechanisms responsible for the formation of the B[e] phenomenon.


\begin{acknowledgements}
This research made use of the NASA Astrophysics Data System (ADS) and of the SIMBAD, VizieR, and 2MASS databases. M.B.F. acknowledges John Monnier, who kindly provided the Keck data. M.B.F. acknowledges financial support from the Programme National de Physique Stellaire (PNPS-France) and the Centre National de la Recherche Scientifique (CNRS-France). M.B.F. also acknowledges Conselho Nacional de Desenvolvimento Cient\'ifico e Tecnol\'ogico (CNPq-Brazil) for the post-doctoral grant. M.K. acknowledges financial support from GA\,AV \v{C}R number KJB300030701.  F.M. and A.M. acknowledge financial support from the Max Planck Institut f\"ur Radioastronomy.
\end{acknowledgements}


\end{document}